\documentstyle[twoside,fleqn,espcrc2,epsf]{article}

\newcommand{\nue}{\mbox{${\nu}_{e}$}}

\newcommand{\num}{\mbox{${\nu}_{\mu}$}}

\newcommand{\nut}{\mbox{${\nu}_{\tau}$}}

\newcommand{\lsim}{\mbox{\raisebox{-1.ex}
{$\stackrel{\textstyle <}{\textstyle \sim}$}}}
\newcommand{\gsim}{\mbox{\raisebox{-1.ex}
{$\stackrel{\textstyle >}{\textstyle \sim}$}}}
\newcommand{\sstt}      {\sin^2 2\theta}
\newcommand{\dms}       {\Delta m^2}
\newcommand{\degree}    {^{\circ}}

\newcommand{\AmS}{{\protect\the\textfont2
  A\kern-.1667em\lower.5ex\hbox{M}\kern-.125emS}}

\title{Atmospheric neutrino results from Super-Kamiokande and Kamiokande \\
 $-~$Evidence for $\num$ oscillations$~-$ }

\author{Takaaki Kajita\address{Kamioka Observatory, Institute for
Cosmic Ray Research, Univ. of Tokyo \\ 
   Higashi-Mozumi, Kamioka-cho, Gifu, 506-1205, Japan } \\
   for the Super-Kamiokande and Kamiokande collaborations }
       
\begin{document}

\begin{abstract}
    New atmospheric neutrino results from Super-Kamiokande are 
presented. Results from Kamiokande on upward going muons 
are also presented. All these data, together with the Kamiokande
atmospheric neutrino data give evidence for neutrino oscillations.
Two flavor $\num \leftrightarrow \nut$ oscillations, with large
sin$^{2} 2 \theta$ and $\Delta$m$^{2}$ in the region of 10$^{-3}$ 
to 10$^{-2}~$eV$^{2}$, explain all these data.
\end{abstract}

\maketitle

\section{Introduction}
   
Cosmic ray interactions in the atmosphere produce 
neutrinos. The prediction of the absolute flux has an
uncertainty of $\pm$20\%. However, the flavor ratio of the 
atmospheric neutrino flux,
$(\nu_{\mu}+\bar{\nu}_{\mu})/(\nu_{e}+\bar{\nu}_{e})$, has been
calculated to an accuracy of better than 5\% in a broad energy
range from 0.1~GeV to higher than 10~GeV.  The
calculated flux ratio has a value of about 2 for energies
$\lsim$1GeV and increases with increasing neutrino energy.  For
neutrino energies higher than a few GeV, the fluxes of upward and
downward going neutrinos are expected to be nearly equal; the 
geomagnetic field effects on atmospheric neutrinos in this
energy range are expected to be small because the primary cosmic
rays that produce these neutrinos have rigidities exceeding the
geomagnetic cutoff rigidity ($\sim 10~GeV/Ze$). 

    The $ \nu_{\mu} / \nu_{e}$ ratio has been measured in deep 
underground experiments by observing final-state leptons produced
by charged-current interactions of neutrinos on nuclei, $\nu + N
\rightarrow l + X$. The measurements are reported as $R~\equiv$   
$(\mu/e)_{data} / (\mu/e)_{MC}$, where $\mu$ and $e$ are the 
number of muon-like ($\mu$-like) and electron-like ($e$-like)
events observed in the detector for both data and Monte Carlo
(MC) simulation. This ratio largely cancels experimental and 
theoretical uncertainties, especially the uncertainty in the 
absolute flux. 
The observed values of $R$ by Kamiokande~{\cite{kam:sub}, \cite{kam:multi}} 
were significantly smaller than unity for both below and above 1~GeV
energy ranges. Consistent results were obtained by
IMB-3($<$1.5GeV)~{\cite{imb:sub}} and Soudan-2~{\cite{soudan:sub}}. 

    Neutrino oscillations have been suggested to explain the
small values of $R$. For a two-neutrino oscillation hypothesis,
the probability for a neutrino produced in a flavor state $a$
to be observed in a flavor state $b$ after traveling a distance 
$L$ through a vacuum is:
\begin{equation}
P_{a \rightarrow b} = \sstt \sin^2{\Big(}\frac{1.27 
\dms(\textrm{eV}^2) L (\textrm{km})}{E_\nu(\textrm{GeV})}\Big{)},
\end{equation}
where $E_\nu$ is the neutrino energy, $\theta$ is the mixing angle
between the flavor eigenstates and the mass eigenstates, and $\dms$ is
the mass-squared difference of the neutrino mass eigenstates. For
detectors near the surface of the Earth, the neutrino flight distance,
and thus the oscillation probability, is a function of the zenith angle of
the neutrino direction. Vertically downward-going neutrinos travel
about 15 km while vertically upward-going neutrinos travel about
13,000 km before interacting in the detector. The broad energy spectrum 
from a few hundred MeV to about 100~GeV
and this range of neutrino flight distances
makes measurements of atmospheric neutrinos sensitive to neutrino
oscillations with $\dms$ down to $10^{-4}$ eV$^2$. 

   The zenith angle dependence of $R$  measured by the 
Kamiokande experiment\cite{kam:multi} at high
energies, together with the small $R$ values, has been cited as 
evidence for neutrino
oscillations. Based on these measurements, 
Kamiokande obtained the allowed parameter regions
of neutrino oscillations. Because of the relatively small
statistics, both $\num \leftrightarrow \nue $ and
  $\num \leftrightarrow \nut$ oscillations were allowed.

    Recently, a long baseline reactor experiment, 
CHOOZ~\cite{chooz}, excluded the $\nu_{\mu} \leftrightarrow 
\nu_{e}$ solution of the atmospheric neutrino problem.   

    The upward-going muons observed in underground detectors are
the products of neutrino interactions in the rock. The mean neutrino
energy of these events are of the order of 100(10)~GeV for 
through-going(stopping) events. These events are  used for
the independent check of the neutrino oscillation analysis of
the lower energy atmospheric neutrino data.

    We present the analyses of atmospheric neutrino events
from Super-Kamiokande. In addition to the events whose vertex
positions are in the fiducial volume of the detector, we present
the upward-going muon results from Kamiokande and Super-Kamiokande.
We observed small values of $R$ and a zenith angle dependent 
deficit of 
$\mu$-like events. While no combination of known uncertainties in the
experimental measurement or prediction of atmospheric neutrino fluxes
is able to explain the data, a two-neutrino oscillation model of
$\num \leftrightarrow \nu_x$, where $\nu_x$ may be $\nut$ or a new,
non-interacting ``sterile'' neutrino, is consistent with the observed 
$R$ values and zenith angle distributions. 
These data and the neutrino oscillation interpretation
were further supported by a small 
(upward stopping muons)/(upward through-going muons) ratio
and zenith-angle dependent deficit of upward through-going muon
flux. From these measurements, we conclude that the atmospheric
neutrino data give evidence for neutrino oscillations.

\section{ Super-Kamiokande detector}

Super-Kamiokande is a cylindrical 50~kton water
Cherenkov detector located at a depth of 2700 meters water equivalent in 
the Kamioka Observatory in Japan. 
The detector consists of an inner detector surrounded by an
outer detector on all sides. The inner and outer detectors are 
optically separated by a pair of opaque sheets.  
11146 50~cm $\phi$ photomultiplier tubes (PMTs), instrumented in 
all surfaces of the inner detector, detect Cherenkov photons
radiated by relativistic charged particles. 1885 20~cm $\phi$
PMTs are instrumented in the outer detector.   The outer detector
is useful for identifying entering cosmic-ray muons and measuring
exiting particles produced by neutrino interactions occurring 
in the inner detector. Pulse height and timing information from
each PMT are recorded and used in the data analysis. The 
trigger threshold for electrons is 5.7~MeV/$c$ at 50\% efficiency.

   For a description of the Kamiokande detector, see Ref.\cite{kamiokande}.

        \section{Fully and partially contained events}

Super-Kamiokande observed a total of 4353 fully-contained (FC)
events and 301 partially-contained (PC) events in a 33.0 kiloton-year
exposure. FC events deposit all of their Cherenkov light in the inner
detector while PC events have exiting tracks which deposit some
Cherenkov light in the outer detector. For the present analyses, the neutrino
interaction vertex was required to have been reconstructed within the
22.5 kiloton fiducial volume, defined to be $>2$~m from the PMT wall.
The number of FC+PC events observed so far in Super-Kamiokande was
about 4 times larger than that in Kamiokande. 

FC events were separated into those with a single visible Cherenkov
ring and those with multiple Cherenkov rings.  For the analysis of FC
events, only single-ring events were used.  Single-ring events were
identified as $e$-like or $\mu$-like based on a likelihood analysis of
light detected around the Cherenkov cone.  The FC events were
separated into ``sub-GeV'' ($E_{vis}<1330$ MeV) and ``multi-GeV''
($E_{vis}>1330$ MeV) samples, where $E_{vis}$ is defined to be the
energy of an electron that would produce the observed amount of
Cherenkov light. $E_{vis}=1330$ MeV corresponds to 
$\sim$1400~MeV$/c$ for muons.

In a full-detector Monte Carlo simulation, 88\% (96\%) of the sub-GeV
$e$-like ($\mu$-like) events were $\nue$ ($\num$) charged-current (CC)
interactions and 84\% (99\%) of the multi-GeV $e$-like ($\mu$-like)
events were $\nue$ ($\num$) CC interactions. PC events
were estimated to be 98\% $\num$ CC interactions; hence,
all PC events were classified as $\mu$-like, and no single-ring
and particle identification requirements were made. 
Table~\ref{tb:evts} summarizes the number of
observed events for both data and Monte Carlo as well as the $R$
values for the sub-GeV and multi-GeV samples. Further details of the 
detector, data selection and event reconstruction used in this analysis are
given in Ref.\cite{sk:subgev,sk:multigev}.

\begin{table}
\begin{center}
\begin{tabular}{lrrr}
\hline\hline
 & Data & Monte Carlo & \\
  \hline
  \multicolumn{3}{l}{sub-GeV} & \\ 
  ~~single-ring   & 2389 & 2622.6 \ \  & \\
  ~~~~$e$-like   & 1231 & 1049.1 \ \  & \\ 
  ~~~~$\mu$-like & 1158 & 1573.6 \ \  & \\ 
  ~~multi-ring    &  911 &  980.7 \ \  & \\ 
  \hline 
  ~~total            & 3300 & 3603.3 \ \  & \\ 
  \hline 
  \multicolumn{4}{l}{$R= 0.63~\pm~0.03(stat.)~\pm~0.05(sys.)$} \\ 
  \hline \hline 
  \multicolumn{3}{l}{multi-GeV} & \\ 
  FC events       &      &       \ \  & \\ 
  ~~single-ring   & 520  & 531.7 \ \  & \\ 
  ~~~~$e$-like   & 290  & 236.0 \ \  & \\ 
  ~~~~$\mu$-like & 230  & 295.7 \ \  & \\ 
  ~~multi-ring    & 533  & 560.1 \ \  & \\ 
  \hline 
  ~~total            & 1053 & 1091.8 \ \  & \\	
  \hline  
  PC events       &      &       \ \  & \\ 
  ~~total(=$\mu -$like) & 301 & 371.6 \ \  & \\
\hline   
  \multicolumn{4}{l}{$R_{FC+PC}= 0.65~\pm~0.05(stat.)~\pm~0.08(sys.)$} \\
\hline\hline
\end{tabular} 
\end{center} 
  \caption{Summary of the sub-GeV, multi-GeV and PC event samples
   observed in 33~kiloton-year exposure of the Super-Kamiokande detector. 
  The data are compared with the Monte Carlo prediction based on the 
  neutrino flux calculation of Ref.\cite{hondaflx}.}
\label{tb:evts} 
\end{table}

\subsection*{Flavor ratio}

${\rm Super-Kamiokande}$ measured significantly small values of $R$ in both 
the sub-GeV
and multi-GeV samples. Several sources of systematic uncertainties in
these measurements were considered\cite{sk:subgev,sk:multigev}. 
For the sub-GeV sample, they were:
5\% from uncertainty in the predicted $\num / \nue $ flux ratio,
3.5\% from the CC neutrino interaction cross
sections and nuclear effects in the H$_{2}$O target, 3\% from the neutral
current cross sections, 2\% from particle identification, 1\% from the
absolute energy calibration, 0.6\% from the vertex fit and fiducial 
volume cut, less than 0.5\% from non-neutrino background events and 
1.5\% from statistical uncertainty in the Monte Carlo. Adding these
errors in quadrature, the total systematic uncertainty was 8\%. 
The systematic uncertainty of $R$ for the multi-GeV sample, obtained 
in a similar way, was 12\%.  Table~\ref{tb:evts} summarizes the 
measured $R$ values in Super-Kamiokande. These results are consistent
with the Kamiokande $R$ values, which were  $0.60~^{+0.06}_{-0.05}
(stat.)~\pm~0.05(sys.)$ for the sub-GeV data and  $0.57~^{+0.08}_{-0.07}
(stat.)~\pm~0.07(sys.)$ for the multi-GeV data\cite{kam:multi}. 
  
    The Super-Kamiokande data
have been analyzed independently by two groups, making the possibility
of significant biases in data selection or event reconstruction
algorithms remote\cite{sk:subgev,sk:multigev}. Given the statistical
error in $R$ especially for the sub-GeV sample, statistical fluctuation
can no longer explain the deviation of $R$ from unity. Assuming that 
the systematic error has the gaussian form,
we estimate the probability that the observed $\mu/e$ ratios could
be due to statistical fluctuation is less than 0.001\% and less than 1\% 
for sub- and multi-GeV samples, respectively.

\begin{table*} [ptb]
\begin{center}
\begin{tabular}{lccccc}
\hline\hline
 & & \multicolumn{2}{c}{Super-Kamiokande} & & 
\multicolumn{1}{c}{Kamiokande} \\
  \hline
 & & MC  & Data & & Data  \\ 
\hline
$e$-like               & & & & & \\
~~Sub-GeV, $<$400MeV/c & & 1.00$\pm$0.04$\pm$0.03	    &
                           1.20$^{+0.11}_{-0.10} \pm{0.03}$ & &
    			   1.29$^{+0.27}_{-0.22}$	    \\	    
~~Sub-GeV, $>$400MeV/c & & 1.02$\pm$0.04$\pm$0.03  	    &
                           1.10$^{+0.11}_{-0.10} \pm{0.03}$ & &
    			   0.76$^{+0.22}_{-0.18}$	    \\	    
~~Multi-GeV            & & 1.01$\pm$0.06$\pm$0.03 	    &
                           0.93$^{+0.13}_{-0.12} \pm{0.02}$ & &
                           1.38$^{+0.39}_{-0.30}$ 	    \\	    
$\mu$-like               & & & & & \\
~~Sub-GeV, $<$400MeV/c & & 1.05$\pm$0.03$\pm$0.02 	    &
                           1.03$^{+0.11}_{-0.10} \pm{0.02}$ & &
    			   1.18$^{+0.31}_{-0.24}$	    \\	    
~~Sub-GeV, $>$400MeV/c & & 1.00$\pm$0.03$\pm$0.02   	    &
                           0.65$^{+0.06}_{-0.05} \pm{0.01}$ & &
    			   1.09$^{+0.22}_{-0.18}$	    \\	    
~~Multi-GeV~(FC+PC)    & & 0.98$\pm$0.03$\pm$0.02 	    &
                           0.54$^{+0.06}_{-0.05} \pm{0.01}$ & &
                           0.58$^{+0.13}_{-0.11}$           \\	    
\hline\hline   
\end{tabular} 
\end{center} 
\caption{Summary of the up/down ratio, $U/D$, for $e$-like and $\mu$-like
events from Super-Kamiokande and Kamiokande. The systematic errors
of the Kamiokande experiment are not shown, but are 
similar to those of Super-Kamiokande.}
\label{up:down} 
\end{table*}
\begin{figure*}[ptb]
\begin{center}
\hspace{0mm} \epsfxsize=10.5cm
 \epsfbox{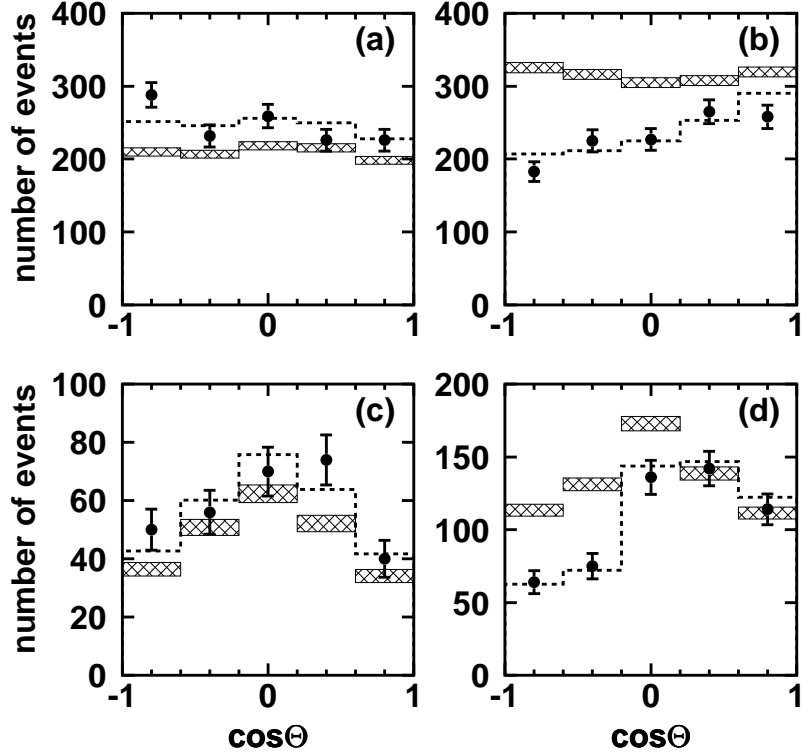}
\vspace{-10mm}
\caption{ Zenith angle distributions observed in Super-Kamiokande 
for; (a)sub-GeV $e$-like, (b)sub-GeV
$\mu$-like, (c)multi-GeV $e$-like and (d) multi-GeV (FC+PC) $\mu$-like 
events. Cos$\Theta =$1 means down-going particles. The histograms with 
shaded error bars show the MC prediction with their statistical errors
for the no neutrino oscillation case.
The dotted histograms shows the Monte Carlo prediction for $\num 
\leftrightarrow \nut$ oscillations with $\sstt =$1 and
$\dms =$2.2$\times$10$^{-3}$~eV$^2$.}
\label{fig:angdist}
\end{center}
\end{figure*}

\subsection*{Zenith angle distribution}
The $\mu$-like data from Super-Kamiokande exhibit a strong up-down 
asymmetry in zenith angle
($\Theta$) while no significant asymmetry is observed in the $e$-like
data, see Figure~\ref{fig:angdist}. For further analyses, we define up-down  
ratio $U/D$ where $U$ is the
number of upward-going events ($-1 <
\cos \Theta < -0.2$) and $D$ is the number of downward-going events 
($0.2 < \cos \Theta < 1$). The ratio is expected to be near unity
independent of flux model for $E_\nu > 1$ GeV, above which effects due
to the Earth's magnetic field on cosmic rays are small. Based on a
comparison of results from our Monte Carlo simulation using
different flux models\cite{gaisserflx,hondaflx} as inputs, treatment
of geomagnetic effects results in an uncertainty of roughly $\pm 0.02
\sim 0.03 $
in the expected $U/D$ values for $e$-like and $\mu$-like sub-GeV events
and less than $\pm 0.02$ for multi-GeV events. These two flux 
calculations do not assume the existence of 1~km mountain 
over the Super-Kamiokande detector. The rock reduces the neutrino flux
because muons are stopped before they can decay in flight. We estimated
the effect of the presence of rock on the flux. $U/D$
 changed about 2\% for the
multi-GeV events. The change was much smaller for the sub-GeV events.   

  Studies of decay
electrons from stopping muons show at most a $\pm 0.6\%$ gain
difference, i.e., measured energy difference, between up-going 
and down-going particles. This gain difference caused 0.9\%, 
and 0.7\% uncertainty in $U/D$ for the multi-GeV 
$e$-like and $\mu $-like 
events. This uncertainty is much smaller for the sub-GeV sample.
A contamination of non-neutrino background such as down-going 
cosmic ray muons could have directional correlation. The maximum 
contribution to the uncertainty in $U/D$ from 
the contamination was 
estimated to be $\pm$1.0\%,  $\pm$0.1\%,  $\pm$0.5\% and $\pm$2.0\% 
for the sub-GeV $e $-like, $\mu $-like, multi-GeV  $e $-like
and $\mu $-like events, respectively. From these studies,
the total systematic uncertainties in $U/D$ for the data 
and MC are summarized in Table~\ref{up:down}.  
In the present data, $U/D$ 
for $e$-like events is consistent with
expectations.  $U/D$ for high momentum $\mu$-like 
events significantly deviates from unity, while  $U/D$  
for low momentum $\mu$-like events is consistent with unity. 
The average angle between
the final state lepton direction and the incoming neutrino direction
is $55\degree$ at $p = 400$ MeV/$c$ and $20\degree$ at 1.5
GeV/$c$.  At the momentum range below 400~MeV/$c$, the possible 
up-down asymmetry of the neutrino flux is largely washed out. 
We have found no detector bias differentiating 
$e$-like and $\mu$-like events that could explain an asymmetry in
$\mu$-like events but not in $e$-like events~\cite{sk:multigev}.

The  $U/D$  value for the multi-GeV FC+PC $\mu$-like
events, $0.54~ ^{+0.06}_{-0.05}~\pm 0.01$ deviates from unity by more 
than 6 standard deviations. This value is also consistent with the
Kamiokande  $U/D$ value for the multi-GeV FC+PC $\mu$-like
events, $0.58~ ^{+0.13}_{-0.12}$. These numbers, which are close to
0.5, suggest a near maximal neutrino mixing. 

\subsection*{Neutrino oscillation analysis}
We have examined the hypotheses of two-flavor $\num \leftrightarrow
\nue$ and $\num \leftrightarrow \nut$ oscillation models using a
$\chi^2$ comparison of the Super-Kamiokande data and Monte Carlo, 
allowing all important
Monte Carlo parameters to vary weighted by their expected 
uncertainties\cite{sk:osc}.
The data were binned by particle type, momentum, and $\cos
\Theta$. A $\chi^2$ is defined as:

{\setlength\arraycolsep{2pt}
\begin{eqnarray}
\chi^2 & = & \sum_{cos\Theta, p}^{70}
\left( \frac{N_{Data} - N_{MC}(\sstt, \dms, \epsilon_{j})}{ \sigma} 
\right)^2 \nonumber \\ 
       &   & + \sum_j \left( \frac{\epsilon_j}{\sigma_j} \right)^2,
\end{eqnarray}}

\noindent
where the sum is over five bins equally spaced in $\cos \Theta$
and seven
momentum bins for both $e$-like events and $\mu$-like plus PC events
(70 bins total).
The statistical error, $\sigma$, accounts for
both data statistics and the weighted Monte Carlo statistics.
$N_{Data}$ is the measured number of events in each
bin. $N_{MC}(\sstt, \dms, \epsilon_{j})$ is the expected number of 
Monte Carlo events and is a function of $\sstt$, $\dms$ and 
$\epsilon_{j}$. $\epsilon_{j}$ are parameters which are related to 
the systematic uncertainties. The parameters (and their uncertainties) 
considered in this analysis are: overall normalization (25\%, but
this was fitted as a free parameter), $E_{\nu}$ spectral index (0.05),
sub-GeV $R$ (8\%), multi-GeV $R$ (12\%), relative normalization
of PC to FC (8\%), $L/E_{\nu}$ (15\%), sub-GeV (2.4\%)
and multi-GeV (2.7\%) up-down ratios. See Ref.\cite{sk:osc} for more
details. 
 
 For $\num
\leftrightarrow \nue$, effects of matter on neutrino propagation
through the Earth were included following
Ref.~\cite{msw}. Due to the small number of events
expected from $\tau$-production (15 to 20 events were expected in the
present FC+PC sample), the effects of $\tau$ appearance and
decay were neglected in simulations of $\num \leftrightarrow
\nut$. A global scan was made on a $(\sstt, \log \dms)$ grid
minimizing $\chi^2$ with respect to uncertainty parameters,
$\epsilon_j$, at each point.

The best-fit to $\num \leftrightarrow \nut$ oscillations,
$\chi^2_{min(phys)} = 65.2 / 67 {\rm ~DOF}$,  was
obtained at $(\sstt, \dms)~=~(1.0,~2.2\times 10^{-3}$~eV$^2$) inside
the physical region ($0 \leq \sstt \leq 1$). The best-fit values of
the Monte Carlo uncertainty parameters were all within their 
expected errors for this point. The global minimum occurred slightly
outside the physical region at $(\sstt,\dms) = 
(1.05, 2.2\times 10^{-3}~$eV$^2), \chi^2_{min(unphys)} = 64.8/67~{\rm
DOF}$. The 90\% C.L. allowed region 
is located at $\chi^2_{min(phys)} + 5.0,$ based on the minimum inside 
the physical region\cite{pdg}. The allowed region is  
shown in Figure~\ref{fig:allowed}. In the region near $\chi^2$ minimum,
the $\chi^2$ distribution is
rather flat and has many local minima so that inside the 90\% interval
the best-fit $\dms$ is not well constrained. The $\chi^2$ increases
rapidly outside of the 90\% C.L. region.  We obtained $\chi^2 =
135/69~{\rm DOF}$, when calculated at $\sstt = 0$, $\dms = 0$
(i.e. assuming no oscillations). 

   The Kamiokande allowed region\cite{kam:multi} obtained by the
contained event analysis is also shown in Figure~\ref{fig:allowed}.
The Super-Kamiokande allowed region favors lower $\dms$ than
that of Kamiokande. However, the allowed regions
from both experiments have a region of overlap. One of the
reasons for the difference in the allowed region is  due
to the zenith angle distribution of the sub-GeV $\mu$-like
events. The  $U/D$ values of the $\mu$-like events from Kamiokande
and Super-Kamiokande are summarized in Table~\ref{up:down}.
It should be noted that Kamiokande  $U/D$ value was consistent 
with unity  for $\mu$-like events in the momentum range of sub-GeV, 
$>$400MeV/{\it c} but Super-Kamiokande observed 
smaller  $U/D$ value for the same sample; a 2.5$\sigma$ difference
in  $U/D$  for this sub-sample. 
This difference results in a difference in the favored $\dms$ region 
in these two experiments, 
since the energy region observing the small  $U/D$  
is directly related to the determination of $\dms$, 

For the test of $\num \leftrightarrow \nue$ oscillations,
the Super-Kamiokande data resulted in a relatively poor fit; 
$\chi^2_{min} = 87.8 / 67
{\rm ~DOF}$, at $(\sstt, \dms) ~=~ (0.93,~3.2\times10^{-3} $
eV$^2)$. The expected  $U/D$ value of the multi-GeV $e$-like 
events for the best-fit 
$\num \leftrightarrow \nue$ oscillation hypothesis, 1.52, differs
from the measured value, $0.93 ~^{+0.13}_{-0.12} ~\pm 0.02$,
by 3.4 standard deviations.
We conclude that the $\num \leftrightarrow \nue$
hypothesis is not favored. 

The zenith angle distributions for the sub- and multi-GeV samples are 
shown in Figure~\ref{fig:angdist}.  The data are compared to the Monte 
Carlo expectation (no oscillations, hatched region) and the best-fit
expectation for $\num \leftrightarrow \nut$ oscillations (dashed
line). The oscillated Monte Carlo well reproduces the zenith angle
distributions of the data. 

In Super-Kamiokande, the oscillation parameters were also estimated by
considering the $R$ measurement and the zenith angle shape separately. 
The 90\%~C.L. allowed regions for each case overlapped at 
$1\times10^{-3} < \dms < 4\times10^{-3}$ eV$^2$ for $\sstt = 1$. 

\begin{figure} [t]
  \epsfxsize=7.5cm
 \epsfbox{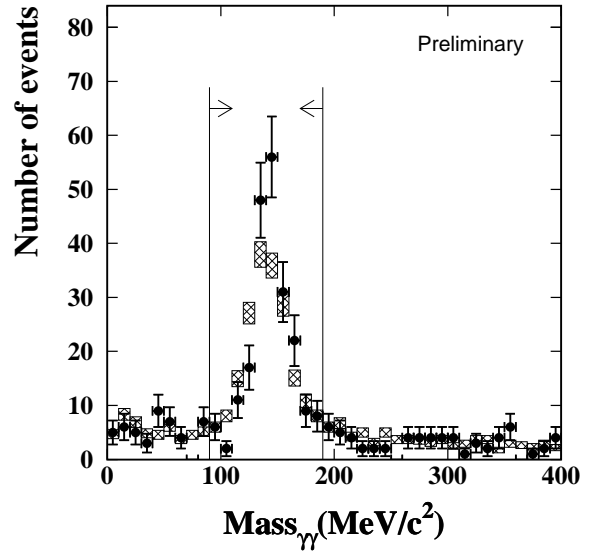}
\vspace{-10mm}
\caption{Invariant mass distribution observed in Super-Kamiokande 
for two ring events with
``$e$-like, $e$-like'' particle-identification results and with no 
$\mu$-decay signal. The histogram with shaded  error bars show
the Monte Carlo prediction with their statistical errors. Cut
region of $\pi^0$-mass is also shown. 
}
\label{pi0}
\end{figure}

As another check of the Super-Kamiokande data, we studied $\pi^0$ events. 
For  $\num \leftrightarrow \nut$ oscillations, the number of 
neutral-current (NC) events should be unchanged by neutrino oscillations.
As a NC sample, we selected ``$\pi^0$''-like events. The selection criteria
were: 2~ring events, both rings should be $e$-like, no $\mu$-decay 
signal and the invariant mass calculated from the charge and
direction of the two rings assuming
two $\gamma$'s should be between 90 and 190MeV/$c^2$. We 
observed 210 $\pi^0$-like events. See Figure~\ref{pi0}, which 
shows the invariant mass distribution for the two ring events.
A clear excess at $\sim$140MeV/$c^2$ is seen. The fraction of NC events
in this sample is estimated to be 82\%.    
For  $\num \leftrightarrow \nut$ oscillations, the number of 
CC $\nue$ events ($\sim e$-like events) should also be unchanged by 
neutrino oscillations. Therefore, the ($\pi^0 / e$) ratio of the
data should agree with the same ratio of the Monte Carlo without 
oscillations for the $\num \leftrightarrow \nut$ case. 
We obtained:  $(\pi^0 / e)_{data} / (\pi^0 /e)_{MC} = 0.93 \pm 0.07
(stat) \pm 0.19(syst)~(Preliminary)$. The result is consistent with
 the $\num \leftrightarrow \nut$ interpretation of the data.

        \section{Upward going muons}

    Energetic atmospheric $\num$'s passing the Earth interact with
rock surrounding the detector and produce muons via CC
interactions. Because of the atmospheric muon background, it is
difficult to select neutrino induced downward going muons. On the
contrary, upward going muons are essentially neutrino origin. 

     Upward going muons can be categorized into two types. One is
``upward through-going muons'' which are the events which enter
into the detector and exit, and the other is ``upward stopping
muons'' which enter the detector and stop in the detector.

\subsection*{Upward through-going muons}

\begin{figure} [t]
  \epsfxsize=8.2cm
\vspace{-7mm}
 \epsfbox{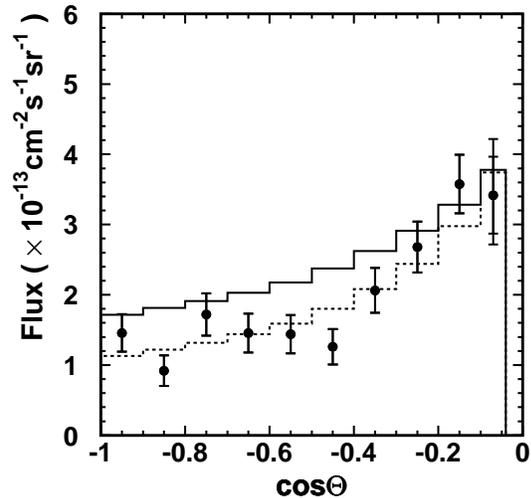}
\vspace{-16mm}
\caption{ Zenith angle distribution of upward through-going muon
flux observed in Kamiokande. Inner(outer) error bars show statistical
(statistical+uncorrelated experimental systematic) errors. The solid
histogram shows the expected flux for the null neutrino oscillation
case. The dashed histogram shows the expected flux for the $\num
\leftrightarrow \nut$ oscillation case with $\sstt =$1.0, 
$\dms =$3.2$\times$10$^{-3}$~eV$^2$ and $\alpha =$1.00. }
\label{kam:upmu:zen}
\end{figure}

The mean energy of neutrinos which produce upward through-going 
muons is about 100GeV.
Kamiokande observed 372 upward going muons during 2456 detector 
live days. The selection criteria were; cos$\Theta =-1 
\sim -0.04$ and the minimum track 
length in the inner detector of 7~meters\cite{kam:upmu}. 
The minimum (mean) energy loss of these muons in the inner
detector is 1.6(3.0)GeV.  The average detection
efficiency was 97\%. With the requirement, cos$\Theta < -0.04$,
the background contamination was negligible. The observed flux of
upward going muons was 1.94$\pm 0.10 (stat.) ^{+0.07}_{-0.06}(sys.)
\times 10^{-13} cm^{-2} sec^{-1} sr^{-1}$.
The expected flux based on the calculated flux of Ref.\cite{gaisserflx}
was 2.46$\pm 0.54(theo.)\times 10^{-13} cm^{-2} sec^{-1} sr^{-1}$. 
Figure~\ref{kam:upmu:zen} shows the zenith angle
distribution of the upward through-going muon flux observed in Kamiokande. 

   In  Super-Kamiokande, 617 upward through-going 
muon events were observed during 537 detector live days\cite{sk:upmu}. 
The selection criteria
were; cos$\Theta <$0, two outer detector cluster corresponding to
the muon entrance and exit points and track length of a muon
inside the inner detector should be longer than 7~m. The minimum
(mean) energy loss of these muons in the inner detector is 
1.6 (6)~GeV. The average 
detection efficiency of these events was estimated to be
$>$99\%. The validity of this efficiency was tested using
the real down-going cosmic-ray muons by assuming the up-down
symmetry of the detector.    

\begin{figure} [t]
  \epsfxsize=7.3cm
 \epsfbox{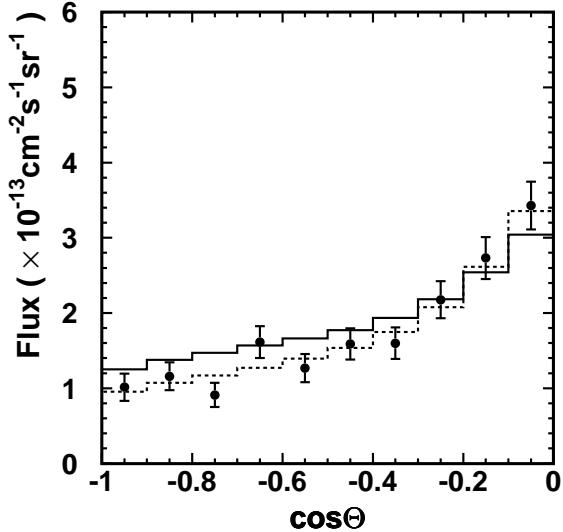}
\vspace{-10mm}
\caption{ Zenith angle distribution of upward through-going muon
flux observed in Super-Kamiokande. Error bars show statistical
+ uncorrelated experimental systematic errors. Estimated background
is subtracted. The solid
histogram shows the expected flux for the null neutrino oscillation
case based on the calculated flux of Ref\cite{hondaflx}. 
The dashed histogram shows the expected flux for the $\num
\leftrightarrow \nut$ oscillation case with $\sstt =$1.0,
$\dms =$2.5$\times$10$^{-3}$~eV$^2$ and $\alpha =$1.12. }
\label{sk:upmu:zen}
\end{figure}

     The number of background events, 4.6, was estimated by 
extrapolating the zenith-angle distribution of cosmic ray muons 
with cos$\Theta =$0$\sim$0.08.
The background events are expected only in the cos$\Theta = -0.1 \sim 0$
bin and are subtracted in further analyses. The observed flux of
upward going muons was 1.75$\pm$0.07$(stat.) \pm$0.09$(sys.)
\times 10^{-13} cm^{-2} sec^{-1} sr^{-1}$.
The expected flux based on the calculated flux of 
Ref.\cite{hondaflx}(\cite{gaisserflx})
was 1.88$\pm 0.42(theo.) \times 10^{-13} cm^{-2} sec^{-1} sr^{-1}$ 
(2.01$\pm 0.45(theo.) \times 10^{-13} cm^{-2} sec^{-1} sr^{-1}$).  
Figure~\ref{sk:upmu:zen} shows the 
zenith-angle distribution of the upward through-going muon flux 
observed in Super-Kamiokande. 

   Because of the difference in the mean track length of muons, the 
observed and predicted fluxes are different between the two experiments. 
However, the
observed/predicted flux ratios are consistent between the two 
experiments within the measurement errors. The measured zenith-angle
distributions in these two experiments have similar shape; both 
experiment observed lower flux near the vertical 
direction compared with the predicted flux distributions. The
$\chi^{2}$ values of the comparison of the shape of the zenith angle
distributions of the data and predictions are 21.3/9~DOF and
18.7/9~DOF for the Kamiokande and the Super-Kamiokande data, 
respectively.    

   Analyses of neutrino oscillations were carried out in these 
experiments. To test the oscillation hypothesis, a $\chi^{2}$
is defined as:
{\setlength\arraycolsep{2pt}
\begin{eqnarray}
\chi^2 & = & \sum_{cos\Theta}^{10}
\left( \frac{\phi_{Data} - \alpha\cdot\phi_{MC}(\sstt, \dms)}
{ \sigma} \right) ^2  \nonumber  \\
    &    & + \left( \frac{\alpha -1}{\sigma_\alpha} \right) ^2,
\end{eqnarray}}

\noindent
where $\sigma$ is the statistical and systematic error in the observed
flux ($\phi_{Data}$) and $\alpha$ and $\sigma_{\alpha}$ are an absolute 
normalization factor of the expected flux($\phi_{MC}(\sstt, \dms)$) 
and its uncertainty, 
respectively.  $\sigma_{\alpha}$ was taken to be $\pm$22\% by
adding the theoretical uncertainty and correlated experimental errors.  
The uncertainty in the absolute neutrino flux ($\pm$20\%) was the 
dominant source of $\sigma_{\alpha}$.
A minimum $\chi^{2}$ was calculated by changing $\alpha$ for each      
(sin$^{2} 2 \theta$, $\Delta m^{2}$). Since the contained data
prefer $\num \leftrightarrow \nut$ oscillations and since the CHOOS
 experiment\cite{chooz} excluded the 
$\num \leftrightarrow \nue$ oscillation parameter region relevant to
the atmospheric neutrino data, 
only $\num \leftrightarrow \nut$ oscillations were tested.  

    The Kamiokande data had $\chi^{2}_{min}$ at (sin$^{2} 2 \theta, 
\Delta m^{2} ) = (1.0, 3.2 \times 10 ^{-3}$~eV$^2$) with $\alpha =$1.00.
The $\chi^{2}_{min}$ value was 12.8/8~DOF. $\chi^{2}_{min}$ for
the Super-Kamiokande data occurred at (sin$^{2} 2 \theta, 
\Delta m^{2} ) = (1.00, 2.5 \times 10 ^{-3}$~eV$^2$) with $\alpha =$1.12.
The $\chi^{2}_{min}$ value was 7.3/8~DOF. The allowed regions did
not change significantly for two assumptions of the neutrino 
flux\cite{hondaflx},\cite{gaisserflx}, 
suggesting that the zenith angle and energy dependences
of the neutrino flux at high energies are understood well. 
 Figure~\ref{fig:allowed}
shows the allowed regions obtained from these upward going muon data.
The allowed regions are larger than those of contained events. 
But they overlap the allowed regions obtained by the contained
data.  

\subsection*{Upward stopping muons}
    Because of the large detector dimension of Super-Kamiokande, 
a substantial fraction
of upward going muons stop in the detector. The mean neutrino energy
of the upward stopping muons is about 10GeV,  which is
substantially lower than that of upward through-going muons.
Therefore, in some neutrino oscillation parameters, it is
expected that an observed (stopping/through-going) flux ratio of 
upward going muons is different from the calculated ratio. 

\begin{figure} [h]
  \epsfxsize=8.2cm
\vspace{-7mm}
 \epsfbox{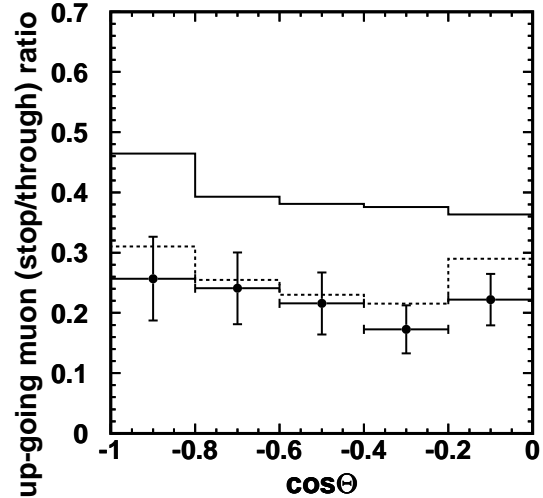}
\vspace{-16mm}
\caption{ Zenith angle distribution of the (stopping/through-going) 
ratio of the upward going muon
flux observed in Super-Kamiokande. Error bars show statistical
+ uncorrelated experimental systematic errors. 
The solid histogram shows the expected ratio for the null
oscillation case. The dashed histogram shows the expected 
ratio for the $\num
\leftrightarrow \nut$ oscillation case with $\sstt =$1.0 and
$\dms = 3\times$10$^{-3}$~eV$^2$. The expected ratio has 
$\pm$14\% correlated uncertainty. }
\label{stop-th:zenith}
\end{figure}

    Super-Kamiokande observed 137 upward stopping muons during
537 detector live days\cite{sk:upmu}. The selection criteria
were similar to those of upward through-going muons except
for a requirement of one outer detector cluster corresponding to the
entrance point. The detection efficiency was estimated to be
99\%. 
 The estimated number of cosmic-ray background events was 13.2. 
It was estimated that these background events were mostly in the 
cos$\Theta = -0.1 \sim 0$ bin. The background events were subtracted 
from this bin for further analyses. We then calculated the  
(stopping/through-going) flux ratio 
($\equiv \Re$). The observed value was $\Re =$0.22$\pm$0.023$(stat.)
 \pm$0.014$(sys.)$, while the predicted value was 
$\Re =$0.39$\pm$0.05$(theo.)$. 
The observed value was substantially smaller than the prediction.
Figure~\ref{stop-th:zenith} shows the zenith angle distribution 
of the  (stopping/through-going) flux ratio. Also shown in the
same figure are the predicted ratios with and without neutrino
oscillations. Clearly, the predicted distribution with
neutrino oscillations agrees well with the data within the 
systematic uncertainty ($\pm$14\%).

\vspace{0mm}
     
The 90\% C.L. allowed region of the neutrino oscillations was
estimated by a $\chi^2$ test: 
{\setlength\arraycolsep{2pt}
\begin{eqnarray}
\chi^2 & = & \sum_{cos\Theta}^{5}
\left( \frac{\Re_{Data} - \beta\cdot\Re_{MC}(\sstt, \dms)}
{ \sigma} \right) ^2   \nonumber   \\
   &    & + \left( \frac{\beta -1}{\sigma_\beta} \right) ^2,
\end{eqnarray}}

\noindent
where $\sigma$ is the experimental (mostly statistical) error in 
the observed ratio ($\Re_{Data}$) and $\beta$ and $\sigma_{\beta}$ are a 
uncertainty factor of the expected ratio ($\Re_{MC}(\sstt, \dms)$)
 and its 1~$\sigma$ error, respectively.  $\sigma_{\alpha}$ was 
taken to be $\pm$14\% by adding the theoretical uncertainty and 
correlated experimental systematic errors.

\begin{figure} [h]
  \epsfxsize=7.5cm
 \epsfbox{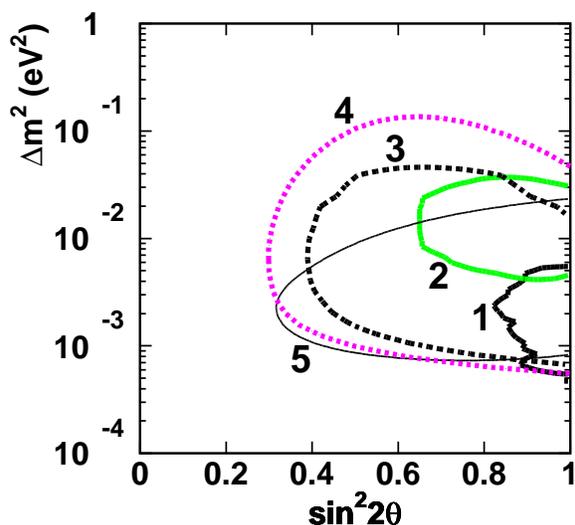}
\vspace{-10mm}
\caption{ The allowed neutrino oscillation parameter regions 
obtained by Kamiokande and Super-Kamiokande. Inside of  
each curve is allowed at 90\% C.L.. The (1)thick-black and 
(2)thick-gray curves show
the regions obtained by contained event analyses from Super-Kamiokande
and Kamiokande, respectively. The (3)black-dotted and (4)gray-dotted 
curves show the 
regions obtained by upward through-going muons from
Super-Kamiokande and Kamiokande, respectively. The (5)thin-black curve 
shows
the region obtained by the (stopping/trough-going) ratio analysis of 
upward going muons from Super-Kamiokande.}
\label{fig:allowed}
\end{figure}

 $\chi^{2}_{min}$  occurred at (sin$^{2} 2 \theta, 
\Delta m^{2} ) = (1.0, 3.4 \times 10 ^{-3}$~eV$^2$).
The $\chi^{2}_{min}$ value was 1.3/3~DOF. Figure~\ref{fig:allowed}
shows the allowed region obtained from this analysis. 
Again, the allowed region is larger than those of contained 
events. But it overlaps the allowed regions obtained by the 
contained and upward through-going muon data.

\section{Conclusions}

    Both the zenith angle distribution of $\mu$-like events
and the ($\mu/e$) values observed in Super-Kamiokande were 
significantly different from the predictions in the absence of 
neutrino oscillations.  While
uncertainties in the flux prediction, cross sections, and experimental
biases are ruled out as explanations of the observations, the present
data are in good agreement with two-flavor $\num
\leftrightarrow \nut$, or  $\num \leftrightarrow \nu_s$,  
oscillations. This conclusion is consistent with
the results from Kamiokande on contained data analysis
and supported by the upward-going muon results from Super-Kamiokande
and Kamiokande. Two experiments, Super-Kamiokande and Kamiokande,
give consistent data and various techniques point to a common parameter
region of neutrino oscillations: $\dms$ should be in the range around
10$^{-3}$ $\sim$ 10$^{-2}$~eV$^2$ and $\sstt \gsim$0.8.
We conclude that the atmospheric neutrino data, especially from 
Super-Kamiokande, give evidence for neutrino oscillations.

\vspace{2mm}
   The Super-Kamiokande experiment is supported by the Japanese Ministry of
Education, Science, Sports and Culture and the United Status Department
of Energy.  The Kamiokande experiment was supported by the Japanese 
Ministry of Education, Science, Sports and Culture.


\begin{thebibliography}{99}

\bibitem{kam:sub}K.S.Hirata {\it et al.}, Phys. Lett. B{\bf 205}
(1988) 416; {\bf 280} (1992) 146.

\bibitem{kam:multi}Y.Fukuda {\it et al.}, Phys. Lett. B {\bf 335}
(1994) 237.

\bibitem{imb:sub}D.Casper {\it et al.}, Phys. Rev. Lett. {\bf 66}
(1991) 2561; R.Becker-Szendy {\it et al.}, Phys. Rev. D {\bf 46} 
(1992) 3720.

\bibitem{soudan:sub}W.W.M.Allison {\it et al.}, Phys. Lett. B {\bf 391}
(1997) 491; E.Peterson, for the Soudan-2 collaboration, in these
Proceedings.

\bibitem{chooz} M.Apollonio {\it et al.}, Phys. Lett. B  {\bf 420} (1998)
397.

\bibitem{kamiokande} K.Nakamura et al., in ``Physics and Astrophysics
of Neutrinos'', Eds., M.Fukugita and A.Suzuki, Springer-Verlag
(1994) p249.

\bibitem{sk:subgev} Y.Fukuda, {\it et al.}, Phys. Lett. B
{\bf 433} (1998) 9.

\bibitem{sk:multigev}Super-Kamiokande Collaboration, 
Y.Fukuda, {\it et al.}, Phys. Lett. B (1998),
accepted for publication, hep-ex/9805006

\bibitem{hondaflx}M.Honda {\it et al.}, Phys. Lett. {\bf B248}
(1990) 193; M.Honda {\it et al.}, Phys. Rev. {\bf D52} (1995) 4985.

\bibitem{gaisserflx}G.Barr {\it et al.}, Phys. Rev. {\bf D39} (1989) 3532;
V.Agrawal, {\it et al.}, Phys. Rev. {\bf D53} (1996) 1313;
T.K.Gaisser and T.Stanev, Proc. 24th Int. Cosmic Ray Conf. 
(Rome) Vol.1 (1995) p694. 

\bibitem{sk:osc}  Y.Fukuda, {\it et al.}, Phys. Rev. Lett. 
{\bf 81} (1998) 1562.

\bibitem{msw} L.Wolfenstein, Phys. Rev. {\bf D17} (1978) 2369; 
S.P.Mikheyev and A.Yu.Smirnov, Sov. J. Nucl. Phys. {\bf 42}
(1985) 1441; S.P.Mikheyev and A.Yu.Smirnov, Nuovo Cim. {\bf C9}
(1986)17.

\bibitem{pdg} Based on a two-dimensional extension of the method
in Review of Particle Properties, Section:
Error and confidence intervals $-$ Bounded physical region:
R.M.Barnett {\it et al.}, Phys. Rev. {\bf D54} (1996) 375.  

\bibitem{kam:upmu}S.Hatakeyama {\it et al.}, 
Phys. Rev. Lett. {\bf 81} (1998) 2016.  

\bibitem{sk:upmu} The Super-Kamiokande collaboration, drafts
in preparation.

\end{thebibliography}
\end{document}